\documentclass[12pt,letterpaper]{article}

\usepackage{float}
\usepackage{times}
\usepackage{graphicx}
\usepackage{amssymb}
\usepackage{amsmath}
\usepackage{multicol}
\usepackage{epstopdf}
\usepackage{placeins}
\usepackage{float}
\usepackage[superscript]{cite} 
\usepackage{lineno}
\usepackage{color}

\newcounter{lastnote}

\title{Phase stabilization by electronic entropy in plutonium
}

\author
{N.~Harrison,$^1$ J.~B.~Betts,$^1$ M.~R.~Wartenbe,$^1$ F.~F.~Balakirev,$^1$ S. Richmond,$^2$\\ M. Jaime,$^1$ P.~H.~Tobash,$^{2\dagger}$\\\\
{\small $^1$Los Alamos National Laboratory, Los Alamos, Mail Stop E536, Los Alamos, NM 87545, USA}\\
{\small $^2$Los Alamos National Laboratory, Los Alamos, Mail Stop E574, Los Alamos, NM 87545, USA}\\
{\footnotesize$^\dagger$To whom correspondence should be sent: nharrison@lanl.gov,}
}

\date{}

\begin{document} 



\baselineskip24pt

\maketitle 

\section*{}
\noindent 

{\bf 
Located at the discontinuity in atomic volume between light and heavy actinides, elemental plutonium (Pu) has an unusually rich phase diagram that includes seven distinct solid state phases and an unusually large 25\% collapse in volume from its $\delta$ phase to its low temperature $\alpha$ phase via a series of structural transitions.\cite{moore1,hecker1,smith1}  Despite considerable advances in our understanding of strong electronic correlations within various structural phases of Pu and other actinides,\cite{georges1,savrasov1,shim1,dai1,wong1,solovyev1,soderlind4,anisimov1,bouchet1,soderlind2} the thermodynamic mechanism responsible for driving the volume collapse has continued to remain a mystery.\cite{lashley2,manley1,jeffries1} Here we utilize the unique sensitivity of magnetostriction\cite{chandrasekhar1} measurements to unstable $f$ electron shells\cite{zieglowski1} to uncover the crucial role played by electronic entropy in stabilizing $\delta$-Pu against volume collapse. We find that in contrast to valence fluctuating rare earths, which typically have a single $f$ electron shell instability whose excitations drive the volume in a single direction in temperature and magnetic field,\cite{zieglowski1,wohlleben1,lawrence1} $\delta$-Pu exhibits two such instabilities whose excitations drive the volume in opposite directions while producing an abundance of entropy at elevated temperatures. The two instabilities imply a near degeneracy between several different configurations of the $5f$ atomic shell,\cite{zwicknagl1,efremov1,eriksson1,wills1,svane1} giving rise to a considerably richer behavior than found in rare earth metals. We use heat capacity measurements to establish a robust thermodynamic connection between the two excitation energies, the atomic volume, and the previously reported excess entropy of $\delta$-Pu at elevated temperatures.\cite{lashley2,manley1,jeffries1} }

\begin{figure}[!!!!!!!tbp]
\begin{center}
\vspace{-2cm}
\includegraphics[angle=-0,width=0.95\linewidth]{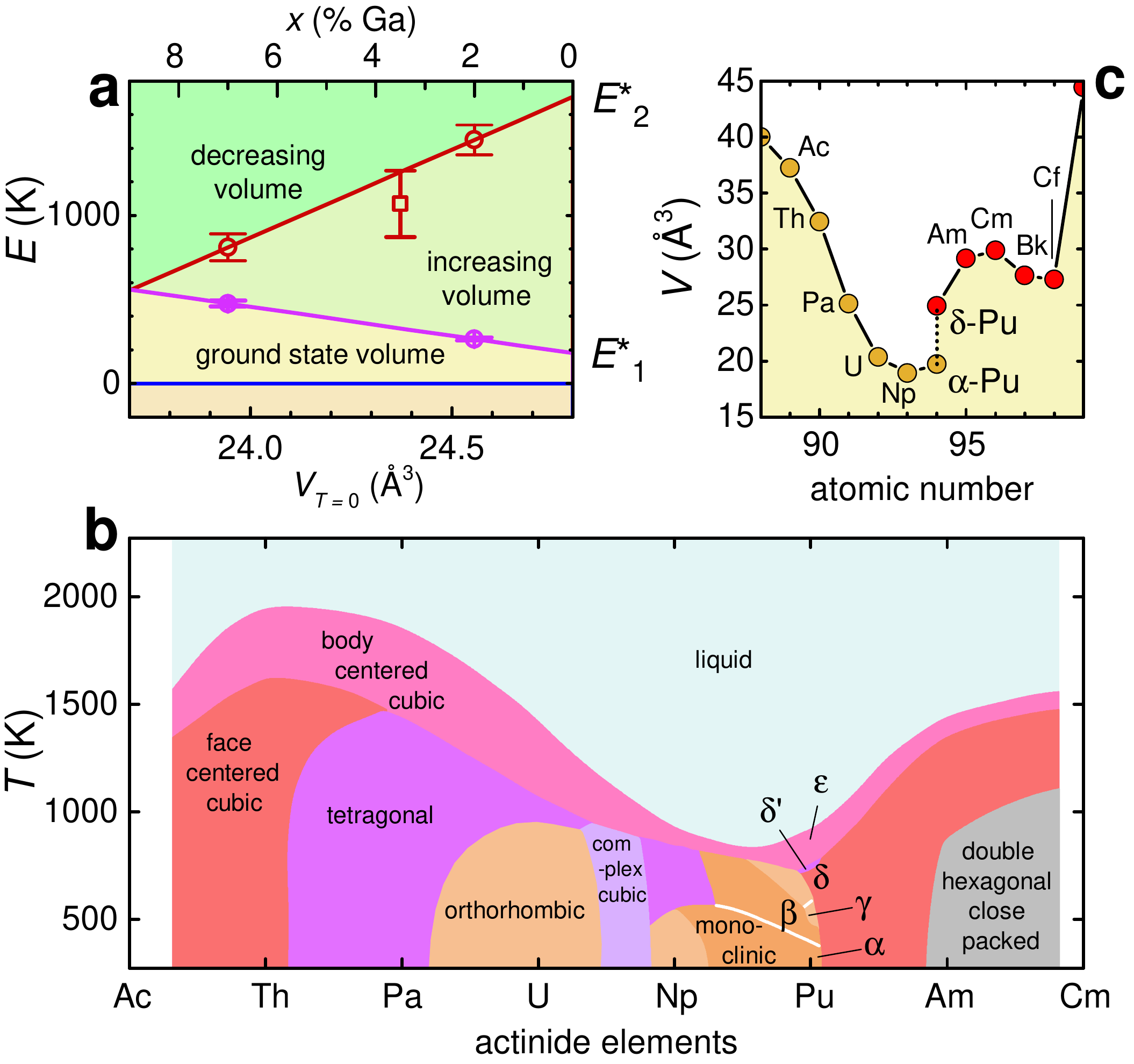}
\vspace{-0.5cm}
\textsf{\caption{{\bf Schematic energies, phases and volumes of Pu.}
{\bf a}, Excitation energies for $\delta$-Pu$_{1-x}$Ga$_x$ for $x=$~2\% and 7\% according to a fit to magnetostriction and thermal expansion measurements (circles), with the lines indicating interpolations assumed during fitting (see Methods). The lower horizontal axis shows the approximate atomic volume of the ground state ($T=0$). Shaded regions indicate the relative volume changes accompanying the excitations. The open square corresponds to the resonance energy observed in neutron scattering measurements,\cite{janoschek1} while error bars indicate its approximate width. {\bf b}, Conceptual connected binary phase diagram of the light actinides, recreated from Refs.,~\cite{smith1,hecker1} with shaded different colored regions representing different types of crystalline structure. The $\alpha$, $\beta$, $\gamma$, $\delta$, $\delta^\prime$ and $\epsilon$ phases of Pu are also indicated. 
{\bf c}, Ground state volume of actinides, as indicated, showing a precipitous drop between (thermally excited or substitutionally-stabilized) $\delta$-Pu and $\alpha$-Pu. 
\label{energies}}}
\end{center}
\end{figure}

\begin{figure}[!!!!!!!htbp]
\begin{center}
\includegraphics[angle=0,width=0.9\linewidth]{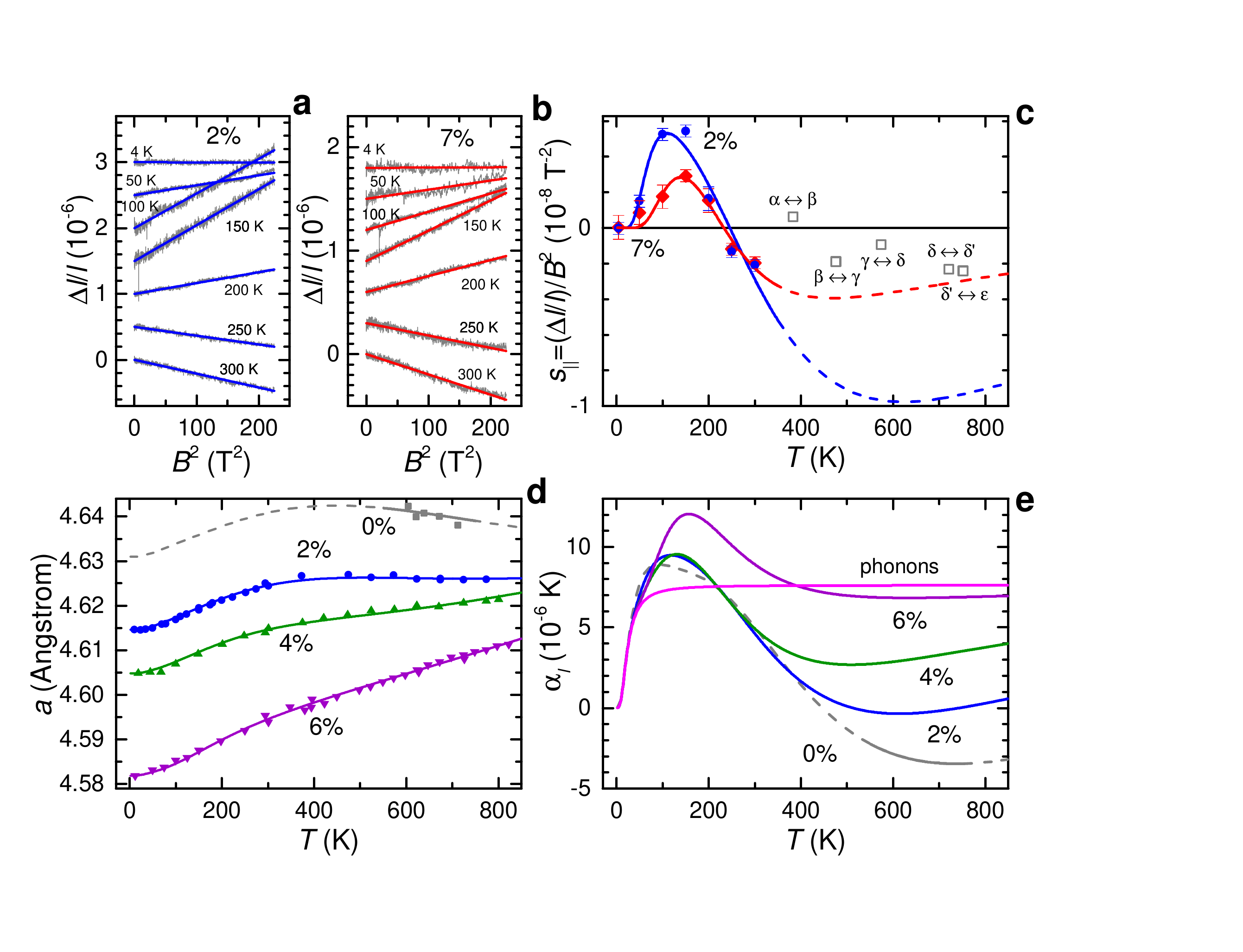}
\textsf{\caption{{\bf Magnetostriction and thermal expansion of Ga-stabilized $\delta$-Pu ($\delta$-Pu$_{1-x}$Ga$_x$)}. {\bf a} Longitudinal magnetostriction (grey) of $x=$~2\% versus $B^2$ at several different temperatures together with quadratic fits (blue lines). {\bf b} Longitudinal magnetostriction of $x=$~7\% versus $B^2$ at several different temperatures together with quadratic fits (red lines). {\bf c} Longitudinal magnetostriction coefficient versus temperature for $x=$~2\% and $x=$~7\%.  Open squares are the magnetovolume coefficient associated with each of the allotropic phase transitions in pure Pu. {\bf d} Lattice parameter $a$ versus $T$ of $x=$~0\%, 2\%, 4\% and 6\% (symbols) from Ref.\cite{lawson1} together with fits (lines)  {\bf e} The thermal expansion, obtained by applying $\alpha=a_{T=0}(\partial a/\partial T)$ to the fitted lines in {\bf d}. The phonon contribution calculated for a Debye temperature of $\theta_{\rm D}=$~100~K\cite{lashley2} is shown for comparison (magenta). The global fit to the data in {\bf c} and {\bf d} are discussed in the Methods. Dashed lines indicate extrapolations. 
}
\label{data}}
\end{center}
\end{figure}

We are able to access the twin excitation energies in plutonium (labelled $E^\ast_1$ and $E^\ast_2$ in Fig.~\ref{energies}a) by way of magnetostriction measurements owing to Ga substitution (in a similar manner to Am substitution in Fig.~\ref{energies}b) affording the stabilization of the $\delta$ phase over a broad span in temperatures and over a range of different volumes.\cite{hecker2,sadigh1} In pure Pu, by contrast, $\delta$-Pu is stable only over a narrow range of high temperatures -- collapsing into significantly lower volume structures upon reducing the temperature (see Figs.~\ref{energies}b and c). We perform magnetostriction measurements on $\delta$-plutonium (see Figs.~\ref{data}a and b) using of an optical fiber Bragg grating technique,\cite{jaime1} which we have adapted for use on encapsulated radiologically toxic  materials (see Methods). The utility of magnetostriction is that, owing to the direct coupling of a magnetic field to magnetic moments, its measurement provides a powerful method for isolating the electronic contribution to the lattice thermodynamics.\cite{chandrasekhar1} While this contribution is vanishingly small in conventional non-magnetic metals, it has been shown to become anomalously large in the vicinity of an $f$-electron shell instability.\cite{kaiser1,thalmeier1,zieglowski1,hafner1,thalmeier2} Furthermore, while changes in the phonon contribution in a magnetic field do generally occur, they occur only in response to a change in the volume that is driven electronically, causing such changes to be a weaker higher order effect.

Figures \ref{data}a and b show measurements of the longitudinal magnetostriction (dilation and contraction along the direction of the magnetic field) of Ga-stabilized polycrystalline plutonium samples of composition $\delta$-Pu$_{1-x}$Ga$_x$ with $x=$~2\% and 7\%. We find that the magnitude of the electronically-driven quadratic-in-magnetic field coefficient of the magnetostriction of $\delta$-Pu (see Fig.~\ref{data}c) falls within the range of values observed in fluctuating valence and Kondo lattice systems.\cite{kaiser1,thalmeier1} However, rather than exhibiting a steep upturn at low temperatures,\cite{zieglowski1,hafner1} as expected for a dominant role played by virtual (or zero point) fluctuations between two or more valence configurations,\cite{savrasov1,shim1,zhu1,janoschek1} the magnetostriction of $\delta$-Pu is observed to vanish at low temperatures. Its behavior closely resembles that of a scenario in which the $f$-electrons condense into a non-magnetic atomic shell configuration, \cite{thalmeier1,barthem1} revealing the electronic excitations to states with different magnetic configurations to be of a predominantly thermally activated nature.

\begin{figure}[!!!!!!!tbp]
\begin{center}
\vspace{-2cm}
\includegraphics[angle=-0,width=0.95\linewidth]{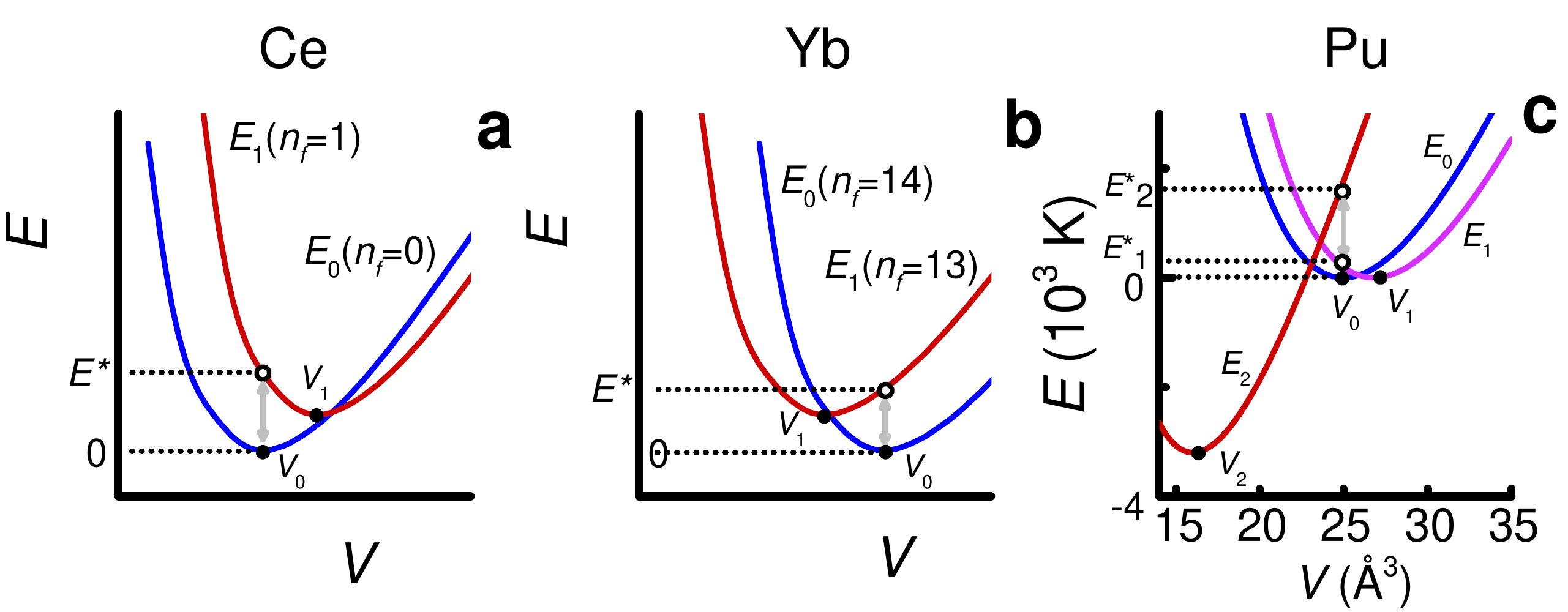}
\vspace{-0.5cm}
\textsf{\caption{{\bf Schematic volume-dependent energies of different electronic configurations.}
{\bf a}, Schematic volume-dependent energies for two configurations of Ce and its compounds, in which $E_0$ has $n_f=0$ $4f$-electrons confined to the atomic core and $E_1$ has $n_f=1$ $4f$-electron confined to the core. The volume-dependent configurational excitation energy $E^\ast_1$ is also indicated. {\bf b}, A similar schematic for Yb. {\bf c}, A possible scheme for $\delta$-Pu that is consistent with $x$-dependent fits to magnetostriction and thermal expansion, consisting of three volume-dependent configurational energies $E_0$, $E_1$ and $E_2$ and two excitation energies $E_1^\ast$ and $E_2^\ast$ (see Methods). In this case, the number of $5f$-electrons confined to the atomic core is unknown.}
\label{levels}}
\end{center}
\end{figure}

When excitations to different electronic configurations occur in $f$-electron systems (e.g. $E_0$ and $E_1$ in Figs.~\ref{levels}a and b), the excited configuration usually has a different number of $f$-electrons confined to the atomic core, causing it to have a different equilibrium atomic volume ($V_1$) and magnetic moment (see Methods).\cite{zieglowski1,wohlleben1} The initial positive increase of the magnetostriction with temperature indicates that the dominant thermal excitations occur between a non-magnetic configuration and a different configuration with both a larger equilibrium atomic size and a larger magnetic moment, as is the case in the majority of $f$-electron systems (see e.g. the illustrated case of Ce in Fig.~\ref{levels}a).\cite{zieglowski1,hafner1,kaiser1,thalmeier1}. However, rather than continuing the same positive sign indefinitely, the sign of the quadratic coefficient of the magnetostriction turns negative beyond $\approx$~200~K (see Fig.~\ref{data}c). A negative sign indicates the onset of thermal excitations into a higher energy electronic configuration with a larger magnetic moment, but whose equilibrium atomic size ($V_2$) is now significantly smaller than that of the other configurations -- as frequently encountered in intermediate valence compounds of Yb (see Fig.~\ref{levels}b).\cite{zieglowski1,hafner1,kaiser1,thalmeier1} The highly non-monotonic temperature-dependence of the magnetostriction in $\delta$-Pu is  indicative of at least three different $f$-electron configurations ($E_0$, $E_1$ and $E_2$) being relevant (shown schematically in Fig.~\ref{levels}c).

Our magnetostriction measurements of $\delta$-Pu are corroborated by thermal expansion measurements,\cite{lawson1} which, while lacking information on magnetic moments of the $f$-electrons, convey more direct information concerning the change in atomic volume between different configurations. The low temperature thermal expansion measurements (see Supplementary Figs.~\ref{lowTthermal}a and b) show the electronic contribution to the thermal expansion from itinerant carriers to be overwhelmed by phonons at temperatures above $\sim$~10~K, as has also been suggested on the basis of heat capacity measurements.\cite{lashley2} 
A low temperature thermal expansion dominated by phonons is further validated by the published temperature-dependent lattice constant data\cite{lawson1} (replotted in Fig.~\ref{data}d). Not until $T\gtrsim$~50~K does a notable departure from the phonon contribution (magenta curve) become apparent, which is shown more clearly in the thermal expansivity (shown in Fig.~\ref{data}e) obtained from a temperature derivative (shown in Fig.~\ref{data}e) of a smooth curve fit to the lattice constant data. The non-phonon contribution to the thermal expansion therefore mirrors the form of the magnetostriction, revealing excitations between electronic configurations to be an equally impactful in both thermodynamic quantities.

\begin{figure}[!!!!!!!htbp]
\begin{center}
\includegraphics[angle=0,angle=0,width=0.9\linewidth]{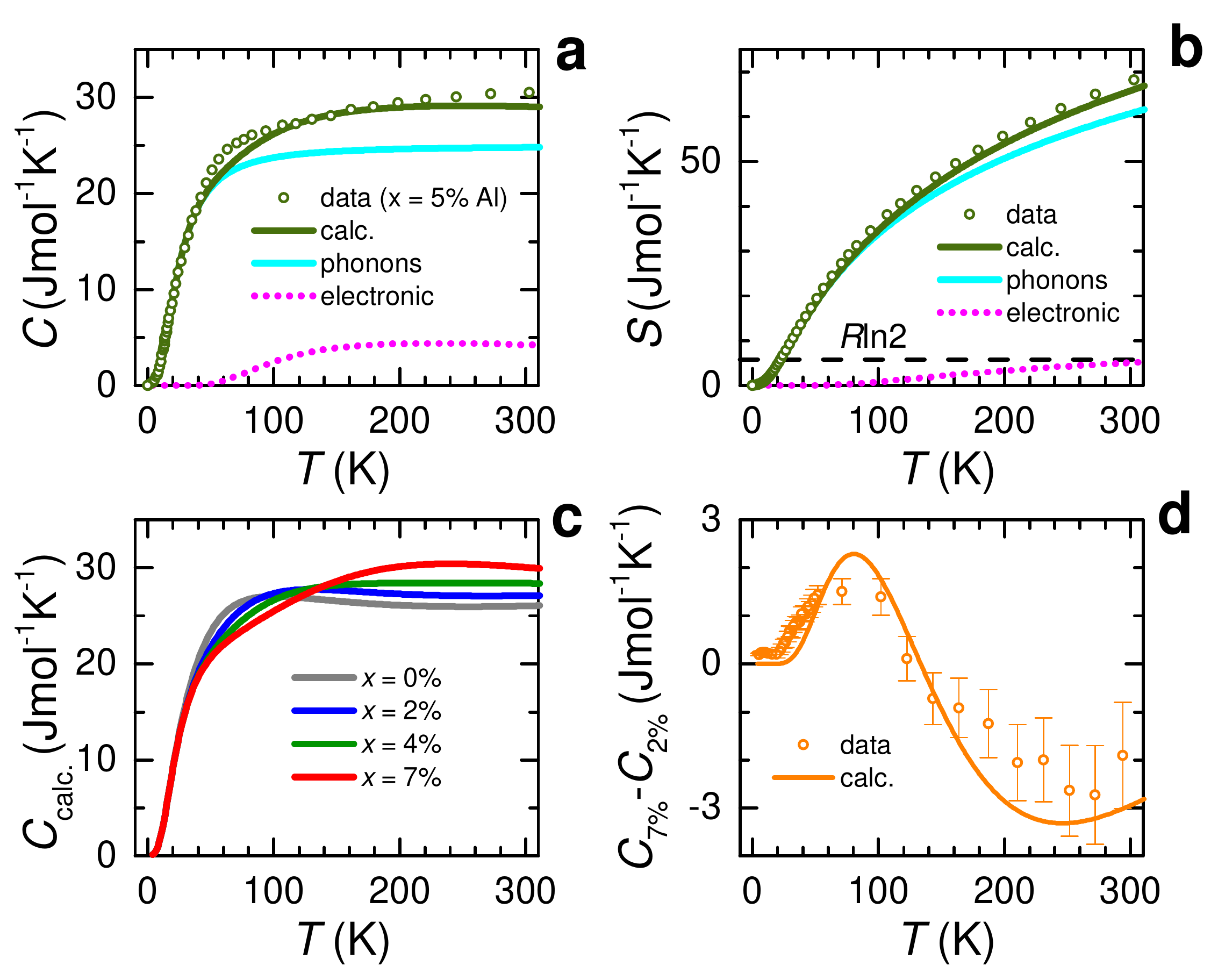}
\textsf{\caption{{\bf Heat capacity and Entropy of Ga-stabilized $\delta$-Pu ($\delta$-Pu$_{1-x}X_x$), where $X=$~Ga or Al}. {\bf a} Calculated heat capacity of  $x=$~5\% (green line) compared against experimental data (for $X=$~Al and $x=$~5\% (circles).\cite{lashley2} The separated calculated phonon (cyan) and electronic (grey) contributions are also shown. We neglect possible differences that may exist between $X=$~Al and Ga. {\bf b} A comparison of the calculated and measured entropy of the same composition, where $S=\int_0^T(C/T^\dagger){\rm d}T^\dagger$. {\bf c} The calculated heat capacity for different values of $x$, as indicated. {\bf d} The calculated difference in heat capacity between $x=$~7\% and $x=$~2\% (line), compared against the experimentally measured difference (circles).
}
\label{entropy}}
\end{center}
\end{figure}

We establish validity of the multiconfigurational picture by showing that the magnetostriction and thermal expansion in Ga-stabilized $\delta$-Pu are fully consistent with a model for the statistical thermodynamics of a multiple level system,\cite{zieglowski1,wohlleben1} and by showing that the model then accurately predicts the temperature and volume-dependence of heat capacity data. When two or more different electronic configurations with different energies coexist at a given value of the atomic volume $V$ (shown schematically in Fig.~\ref{levels}), their relative occupations can be described by a partition function $Z_{\rm el}$ (see Methods), which produces an electronic contribution to the free energy of the form $F_{\rm el}=-k_{\rm B}TN\ln Z_{\rm el}$. Thermodynamic quantities, such as the quadratic-in-field magnetostriction coefficient $s_\nu$, thermal expansion $\alpha_\nu$ and heat capacity $C_v$, are then given by second derivatives ($s_\nu=-\frac{\kappa_0}{B}\frac{\partial^2F}{\partial\nu\partial B}$, $\alpha_\nu=-\kappa_0\frac{\partial^2F}{\partial \nu\partial T}$ and $C_p\approx C_v=-T\frac{\partial^2F}{\partial T^2}$) of the total free energy $F=F_{\rm el}+F_{\rm ph}$, where $F_{\rm ph}$ is the contribution from phonons (see Methods).\cite{morse1,lawson1} To demonstrate validity of the multiconfigurational state we {\it first} perform a simultaneous fit of $F(T,B,x)$ to two measured thermodynamic quantities, namely the magnetostriction and thermal expansion (lines in Figs.~\ref{data}c and d), using a single set of parameters. We {\it then} show that the same form for $F(T,B,x)$ successfully predicts a third thermodynamic quantity, namely the heat capacity (see Fig.~\ref{entropy}a). On computing the heat capacity in Fig.~\ref{entropy}a using the form of the free energy extracted from magnetostriction and thermal expansion measurements, we find multicofigurational excitations to add $\sim$~5~Jmol$^{-1}$K$^{-1}$ to the heat capacity at $T\gtrsim$~100~K, bringing it into close agreement with the published experimental curve.\cite{lashley2} Multiconfigurational electronic excitations therefore produce the largest contribution to the heat capacity and entropy after phonons (see Figs.~\ref{entropy}a and b), with the characteristic energy $E^\ast_1$ dominating at temperatures between $\approx$~100 and 300~K and $E^\ast_2$ coming in at higher temperatures (see Fig.~\ref{levels}c). Importantly, the entropy associated with the electronic excitations (see Fig.~\ref{entropy}b) is more than sufficient to account for the $\approx$~0.8~$\times R\ln2$ excess entropy previously identified as favoring the stabilization of $\delta$-Pu over $\alpha$-Pu at high temperatures (where $R=k_{\rm B}N_{\rm A}$ and $N_{\rm A}$ is Avogadro's number).\cite{jeffries1,manley1}

A particularly striking finding is that the excitation energies $E^\ast_1$ and $E^\ast_2$ change rapidly as a function of the Ga content $x$ (plotted in Fig.~\ref{energies}a), and in opposite directions. One predicted consequence of their rapid change with $x$ is that the heat capacity is expected to become strongly dependent on the Ga composition (see Fig.~\ref{entropy}c), thus providing a means for the extreme dependences of $E^\ast_1$ and $E^\ast_2$ on $x$ to be robustly verified by experiment.  To confirm that the extreme sensitivity of $E^\ast_1$ and $E^\ast_2$ to $x$ and $V$ is a genuine effect, we calculate the heat capacity as a function of $T$ at different values of $x$ (see Fig.~\ref{entropy}c) and compare it against an independent set of $x$-dependent heat capacity measurements (raw data in Supplementary Fig.~\ref{tobash}). On taking the difference between calculated heat capacity for $x=$~2\% and 7\% (in Fig.~\ref{entropy}c), we find that it {\it indeed} accurately predicts the difference in heat capacity observed experimentally (shown in Fig.~\ref{entropy}d), including both the absolute magnitude of the difference and the existence of a sign change in the difference at $\approx$~130~K. Since the primary effect of Ga substitution is to reduce the volume $V$ of $\delta$-Pu (the ground state atomic volume of $\delta$-Pu$_{1-x}$Ga$_x$ being shown on the lower horizontal axis of Fig.~\ref{energies}a),\cite{hecker2} we attribute the opposite variations of $E^\ast_1$ and $E^\ast_2$ with $x$ to their sensitivity to volume, illustrated in Fig.~\ref{levels}c.

We have therefore uncovered two previously hidden electronic energy scales giving rise to significant entropy in excess of the Dulong-Petit value\cite{petit1} of $\approx$~25~Jmol$^{-1}$K$^{-1}$ at temperatures above the Debye temperature in plutonium. The strong variations of $E^\ast_1$ and $E^\ast_2$ with $x$ shed light on the long unresolved questions\cite{hecker1} of why the volume collapse occurs and why it is inhibited by Ga substitution.\cite{hecker2} A likely energetic motivation for the volume collapse at low temperatures is provided by the steep decline in $E^\ast_2$ with decreasing volume (see Fig.~\ref{energies}a), which suggests an equilibrium volume ($V_2$) for $E_2$, similar to that found by density functional theory,\cite{bouchet1} that is lower in energy than that of $E_0$ in Fig.~\ref{levels}c. In practice, the volume collapse (which takes place via a series of steps in pure Pu)\cite{hecker1,sadigh1} is accompanied by additional translational symmetry breaking into the $\alpha$ phase (or $\alpha^\prime$ phase in the alloys), which give rise to an energy barrier between the different volume phases.\cite{jeffries1} In pure $\delta$-Pu, $E^\ast_2$ is too high ($\approx$~1700~K according to an extrapolation to $x=$~0\% in Fig.~\ref{energies}a) to supply sufficient entropy to $\delta$-Pu for it to remain stable over an extended range of temperatures, resulting in its ultimate collapse into the $\alpha$ phase. Conversely, in heavily substituted $\delta$-Pu$_{1-x}$Ga$_x$, $E^\ast_1$ and $E^\ast_2$ are both sufficiently low to afford the $\delta$ phase in the alloy a significantly elevated entropy relative to that in pure plutonium, as evidenced by the higher electronic heat capacity above $\approx$~130~K of heavily Ga-stabilized $\delta$ plutonium relative to its Ga-reduced counterpart. This additional entropy thus provides $\delta$-Pu$_{1-x}$Ga$_x$ with additional protection against collapsing into the $\alpha$ phase.\cite{hecker2,sadigh1,jeffries1}

The strong interplay between the excitation energies and $V$ is further demonstrated by the similarity in behavior of the magnetostriction to the transitional magnetovolume coefficient associated with each of the crystallographic phase transitions in plutonium.\cite{lashley1} We estimate the transitional magnetovolume coefficient, $s_\|^{\rm mv}\approx\frac{\mu_0\kappa_0}{6}\Delta\chi/\Delta\nu$ (open squares in Fig.~\ref{data}c), of pure Pu from the ratio of the previously measured jump $\Delta\chi$ in the susceptibility to the jump $\Delta\nu$ in volume dilation (plotted in Supplementary Fig.~\ref{extrastuff}) at each of the phase transitions. We find the magnetovolume and magnetostriction coefficients to be of similar magnitude and to exhibit similar positive-to-negative trends with increasing temperature, implying that $E^\ast_1$ and $E^\ast_2$ respond similarly to reductions in $V$ caused by phase transitions as they do to reductions in $V$ caused by Ga substitution.

Our findings shed new light on the electronic structure of plutonium and its relation to other actinides and to rare earths. While energy scales of comparable magnitude to the larger excitation energy ($E^\ast_2$ in Fig.~\ref{energies}a) have been inferred from neutron scattering experiments (for $x=$~3.5\%\cite{janoschek1} plotted in Fig.~\ref{energies}a) and from fits made {\it solely} to the thermal expansion,\cite{lawson1} their origin have remained controversial\cite{janoschek1,migliori1} while their volume-dependences have remained unknown. Electronic structure calculations have shown that Pu is able to exist in a larger number of near degenerate configurations than most other $f$-electron systems,\cite{eriksson1,svane1,shick1,soderlind1,wills1} with each having a different number of $5f$-electrons confined or localized within the atomic core and different values of the equilibrium atomic volume, thereby providing a likely origin for $E_0$, $E_1$ and $E_2$ (shown schematically in Fig.~\ref{levels}c). One implication of the demonstrated validity of a multiconfigurational partition function in Pu is that virtual valence fluctuations\cite{savrasov1,shim1,zhu1} appear to be relatively unimportant for understanding the statistical thermodynamics of plutonium (see Methods). On incorporating the valence fluctuation temperature $T_{\rm fl}$ phenomenologically into $F(T,B,x)$,\cite{zieglowski1,wohlleben1} we find that $T_{\rm fl}\lesssim$~50~K (see Methods), which is more than an order of magnitude smaller than the effective Kondo temperature of $T_{\rm K}\sim$~10$^3$~K suggested by contemporary electronic structure methods.\cite{savrasov1,shim1,zhu1} A smaller energy scale can be more easily reconciled with the conventional linear-in-$T$ Sommerfeld contribution that persists to only $\sim$~20~K in heat capacity experiments,\cite{lashley2} which justifies our being able to neglect its contribution to the entropy, and also places Pu's largest volume phase, $\delta$-Pu, in a thermodynamically equivalent regime to the largest volume phase of cerium, $\gamma$-Ce.\cite{lawrence1} In the valence fluctuating rare earth systems,\cite{lawrence1,immer1} electronic entropy has similarly been reported as the crucial factor determining the stabilization of their high temperature phases.\cite{drymiotis1,dzero1,amadon1} The essential difference in Pu is that the twin instabilities, $E^\ast_1$ and $E^\ast_2$, provide much more entropy to the system, enabling it to undergo larger volume changes and more numerous structural transitions.

\newpage


%
%



\bibliographystyle{naturemag}



%




\section{Acknowledgements}
The work was performed under the Los Alamos National Laboratory LDRD program: project ``20180025DR.'' Measurements were performed at the National High Magnetic Field Laboratory, which is supported by the National Science Foundation, Florida State and the Department of Energy.

\section{Author Contributions}
N. H., J. B. B., M. R. W. and M. J. performed the measurements. P. H. T. and S. R. prepared and mounted the samples. M. J., J. B. B., M. R. W. and F. F. B. developed the experimental apparatus. N. H. performed the modeling. N. H., P. H. T. and M. J. wrote the manuscript. P. H. T. arranged all of the sample transportation logistics.

\section{Competing interests} 
There are no competing interests.

\section{Materials \& Correspondence}
Please send correspondence and meterials requests to N.~Harrison; email:nharrison@lanl.gov.


\clearpage

\section{Methods}

\subsection{Magnetostriction measurements}
Variations $\Delta l_\|$ in the sample length $l_\|$ along the direction of the magnetic field are measured either upon sweeping the temperature at zero magnetic field or on sweeping the magnetic field up to 15~T and back to zero at fixed temperature, for both polarities of the magnetic field. The measurements are made using the fiber Bragg grating method,\cite{jaime1} in which we record the spectral information on the light reflected by 1 and 2~mm long Bragg gratings inscribed in the core of a 125~$\mu$m telecom-type optical fiber. A flat face of a sample is attached to a single grating on its own fiber using cyanoacrylate glue. One or two `empty' gratings on the same fiber provide a means for compensating for the temperature-dependence of the diffraction index of the fiber in the absence of a sample. 

Multiple fibers are fed through stainless steel capillary tubes into a brass can that forms the body of the sample primary encapsulation. Using this method, multiple samples can be co-encapsulated, while a steel hepa filter enables $^4$He gas or liquid to circulate. The fibers holding the samples are anchored to a metallic block for thermalization, made of non-magnetic stainless steel in the case of samples 1 and 2 and copper in the case of samples 3 and 4. Thermometers are also anchored to the metallic block inside the can. The brass can is then mounted on the end of a probe inside a secondary containment containing either vacuum or $^4$He, which can be pumped through a high through-put hepa filter situated on the pumping line. The secondary containment, which has its own thermometer, is then placed inside a variable temperature insert (VTI) that itself goes inside the bore of a 15~T superconducting magnet. 

The temperature is controlled via the VTI by using a heater and also, when necessary, using a secondary heater on the secondary containment. Using this arrangement, the temperature can be stabilized to $\approx$~50~mK, with a small thermal drift occurring over timescales of order several hours. To eliminate the effect of thermal drift during magnetostriction measurements, up and down sweeps of the magnetic field are averaged and the temperature adjusted accordingly. Negative and positive sweeps of the magnetic fields are also compared to ensure reproducibility of the result.

\begin{figure}[!!!!!!!htbp]
\begin{center}
\includegraphics[angle=0,width=0.45\linewidth]{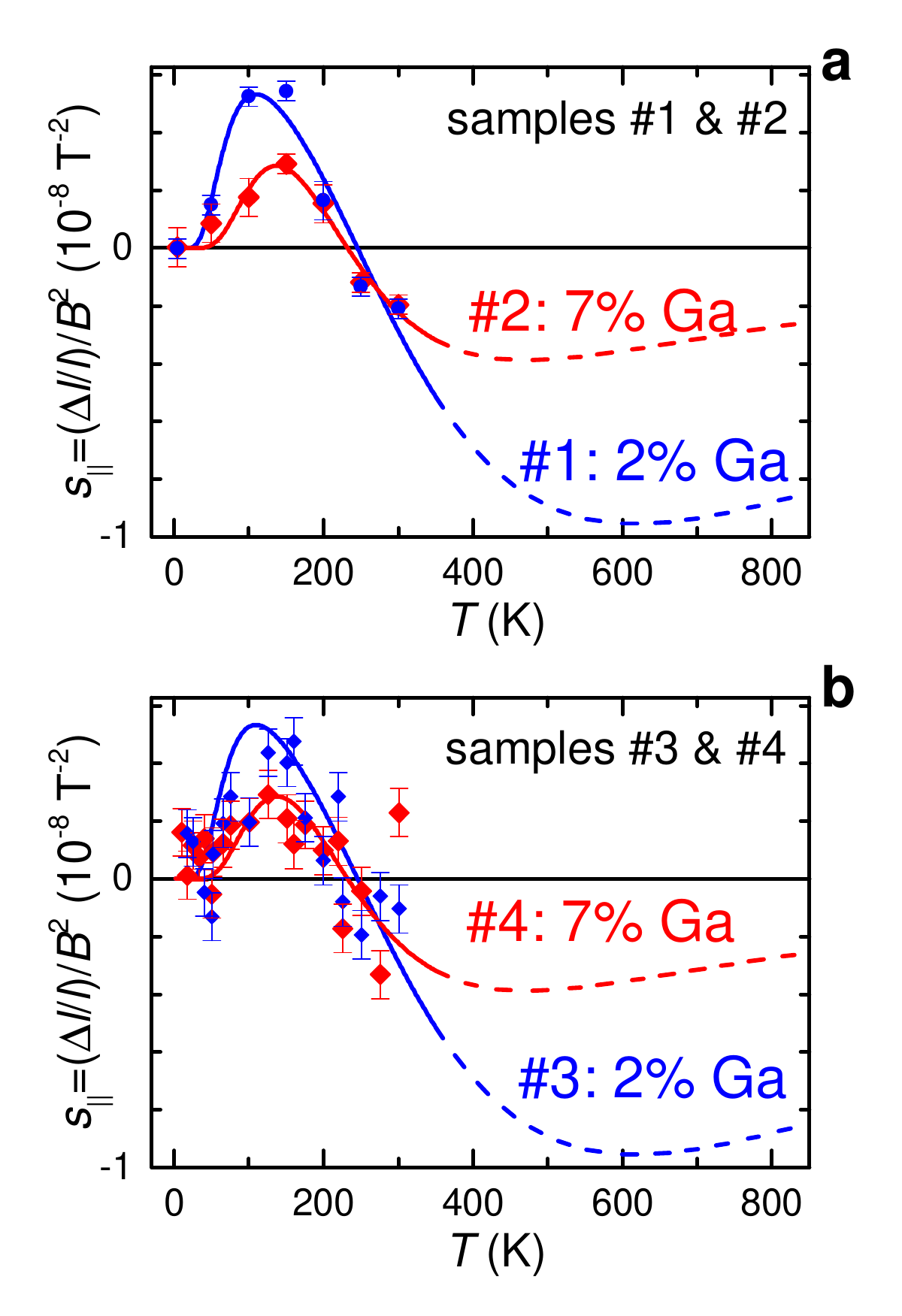}
\textsf{\caption{{\bf Coefficients of the quadratic magnetostriction coefficient versus $T$ for four different $\delta$-Pu$_{1-x}$Ga$_x$ samples, as indicated}. 
}
\label{otherdata}}
\end{center}
\end{figure}

Figure \ref{otherdata} shows the quadratic magnetostriction coefficient as a function of $T$ on four different samples. A lower signal-to-noise ratio is observed in the case of samples 1 and 2, which we therefore use for performing fits. Samples 3 and 4 are found to have magnetostriction coefficients that are consistent with the fits to samples 1 and 2. Error bars are estimated after repeating the magnetostriction measurements at the same temperature, often with a different polarity of the magnetic field.

\subsection{Sample preparation details for magnetostriction measurements} 
Several polycrystalline samples of  $\delta$-Pu$_{1-x}$Ga$_x$ with $x=$~2\% and 7\% were prepared in the form of plates of a few millimeters with masses ranging between 16 and 40 mg, and are annealed prior to mounting for magnetostriction measurements. The 2\% $\delta$-Pu gallium-stabilized samples are homogenized at 450~$^\circ$C while the 7\% $\delta$-Pu gallium-stabilized samples are homogenized at 525~$^\circ$C. Samples 1 and 2, measured in the main text, have Ga compositions of 2\% and 7\% with masses of 16.2 mg and 30.7 mg, respectively and dimensions on the order of $\sim$~1~mm~$\times$~4~mm with a thickness of 150~$\mu$m. The samples were lightly polished prior to gluing onto the fibers in order to remove any surface oxidation. The glue also has the effect of protecting the measured flat surfaces of the samples against oxidation during their loading into the VTI.

For the $x=$~2\% sample, the sample length is observed to drift slowly in time when the temperature is set close to $\approx$~150~K as a consequence of the partial and gradual transformation of $\delta$-Pu into the $\alpha^\prime$ phase of plutonium. Here, the $\alpha^\prime$ phase in Ga substituted Pu has the same structure as the $\alpha$ phase in pure Pu. The total change experienced during the course of the stabilization at $\approx$~150~K is $\Delta l/l\approx-$~0.15~\%, which, given the smaller atomic volume of $\alpha$-Pu, corresponds to 2.3\% of the sample (by volume) transforming. No similar transformation is observed on measuring the $x=$~7\% sample.  

\subsection{Thermodynamics} 
The coefficients of thermal expansion and volume magnetostriction are given by\cite{thalmeier2}
\begin{equation}\label{derivatives}
\frac{\partial\nu}{\partial T}=\alpha_\nu=-K_0^{-1}\frac{\partial^2F}{\partial \nu\partial T}~~~{\rm and}~~~s_\nu B=\lambda_\nu=-K_0^{-1}\frac{\partial^2F}{\partial\nu\partial B},
\end{equation}
respectively, where $F$ is the free energy, $T$ is temperature, $B$ is the magnetic field, $K_0$ is the bulk modulus and $\nu=(\Delta V/V_0)$ is the volume expansion (or contraction). In the absence of broken time reversal symmetry (e.g. a ferromagnetic and some types of non-collinear antiferromagnetic ground state),\cite{jaime2017} $\lambda_\nu(B)$ is linear in magnetic field, in which case the volume increases quadratically with field with the coefficient $s_\nu=\lambda_\nu/B$.

We assume the free energy to be the sum $F=F_{\rm el}+F_{\rm ph}$ of electronic and phonon contributions. The phonon contribution is given by\cite{lawson1,morse1}
\begin{equation}\label{phonons}
F_{\rm ph}=Nk_{\rm B}T\Bigg[\frac{8}{9}\frac{\Theta_{\rm D}(1+\nu)^{-\gamma}}{T}+3\ln\bigg[1-{\rm e}^{-\frac{\Theta_{\rm D}(1+\nu)^{-\gamma}}{T}}\bigg]-{\rm D}\bigg(\frac{\Theta_{\rm D}(1+\nu)^{-\gamma}}{T}\bigg)\Bigg]
\end{equation}
where $N$ is the atomic density (inverse of the unit cell volume) and ${\rm D}(x)$ is the Debye function and we have used $\Theta_{\rm D}=$~100~K,\cite{lashley2} while $\gamma\approx$~0.5.\cite{lawson2} 

For systems with multiple electronic configurations that have the potential to coexist,\cite{hirst1} we assume that each configuration $i$ has its own unique energy $E_i(\nu)$ that depends on $\nu$ in the manner illustrated in Fig.~\ref{levels}, and as predicted to be the case in plutonium.\cite{eriksson1,svane1} The multiconfigurational partition function can then be written in the form
\begin{equation}\label{partitionfunction}
Z_{\rm el}=\sum_{i=0,1,2}\sum_{\sigma=\pm\frac{1}{2}} e^{-\frac{k_{\rm B}E_i(\nu)+2\sigma\mu_i^\ast B}{k_{\rm B}T^\prime}}
\end{equation}
where $\sigma$ refers to spin $\frac{1}{2}$ pseudospins with $g=2$. Note that the summation is made over configurations that have different functional forms for $E_i(\nu)$, but are always at the same volume $V$. $V_i$ refers to the `equilibrium volume' at which a given configuration would be located, were it to have the lowest energy at $T=0$. The multiconfigurational electronic contribution to the free energy is given by 
\begin{equation}\label{electronicfreeenergy}
F_{\rm el}=-N{k_{\rm B}T^\prime}\ln Z_{\rm el}.
\end{equation}

The multiple configurations consist of states in which different numbers $n_f$ of $f$-electrons are confined to the atomic core, or different crystal electric field levels in which the same number of $f$-electrons are confined to the atomic core.\cite{zieglowski1,wohlleben1} However, the latter are typically more closely spaced in energy and volume. To minimize the number of fitting parameters, we assume an effective moment $\mu_i^\ast$, which refers either to that of the lowest crystal electric field level or an average over two or more occupied levels for a given value of $n_f$. The Van Vleck contribution can also add to $\mu_i^\ast$, as this is known to vary as a function of $n_f$.

It has been shown that virtual valence fluctuations can be phenomenologically modeled\cite{wohlleben1,zieglowski1} by introducing an effective valence fluctuation temperature $T_{\rm fl}$, such that $T^\prime=(T^m+T_{\rm fl}^m)^{\frac{1}{m}}$ where $m=$~1 or 2 in Equations (\ref{partitionfunction}) and (\ref{electronicfreeenergy}). In typical rare earth intermediate valence systems, there are only two relevant configurations states that need to be taken into consideration (e.g. Figs.\ref{levels}a and b). However, fits to the plutonium magnetostriction and lattice parameter data (below) find that a best fit is obtained for $T_{\rm fl}=$~0~K, rendering the inclusion of valence fluctuations somewhat unnecessary (see Table~\ref{table1}). The largest value of $T_{\rm fl}$ compatible with experimental data is $\approx$~50~K (see below), and only for $m=2$. 

\subsection{Numerical simulations and fitting procedure} 
To facilitate fitting to experimental data,\cite{zieglowski1,wohlleben1} we first differentiate $F$ with respect to the $\nu$ to obtain
\begin{equation}\label{fractionalvolume}
\nu(T,B)\approx\int_0^T\alpha_\nu(T^\prime,B){\rm d}T^\prime=-K_0^{-1}\frac{\partial F}{\partial\nu}\bigg|_p,
\end{equation}
where, here, $K_0$ refers the bulk modulus of the ground state configuration. For the phonon contribution, we proceed to calculate its contribution ($\nu$) numerically. For the electronic contribution, by contrast, differentiation yields the conveniently trivial result
\begin{equation}\label{electronicvolume}
\nu_{\rm el}(T,B)\approx\frac{1}{{Z_{\rm el}}^\ast}\sum_{i=0,1,2}\sum_{\sigma=\pm\frac{1}{2}} \nu_i^\ast e^{-\frac{k_{\rm B}E_i^\ast+2\sigma\mu_i^\ast B}{k_{\rm B}T^\prime}}, 
\end{equation}
where
\begin{equation}\label{newpartition}
{Z_{\rm el}}^\ast=\sum_{i=0,1,2}\sum_{\sigma=\pm\frac{1}{2}}e^{-\frac{k_{\rm B}E_i^\ast+2\sigma\mu_i^\ast B}{k_{\rm B}T^\prime}}.
\end{equation}
Since the overall extent of the volume expansion in Fig.~\ref{data}d is $\nu\lesssim$~0.6\% for $x=$~2\% and $\nu\lesssim$~2\% for $x=$~6\% samples, we have simplified the fitting procedure by setting $\nu$ to zero on the right-hand-side after differentiating. Following through with this approximation amounts to neglect of a possible 4~K temperature-dependent shift in $E_i$ for $x=$~2\% and a possible 30~K shift for $x=$~7\%. The changes in $E_i$ with $T$ are significantly less than the experimental uncertainty for $x=$~2\% and comparable to the experimental uncertainty for $x=$~7\% (typical error bars listed in Table~\ref{table1}). By comparison, volume changes induced by a magnetic field are only of order 1~ppm. Setting $\nu=$~0 on the right-hand-side simplifies the fitting procedure by allowing us to adopt effective parameters: $\nu_i^\ast=Nk_{\rm B}K_0^{-1}\times\frac{\partial E_i}{\partial\nu}$ and $E_i^\ast=E_i(\nu=0)$ in Equations (\ref{electronicvolume}) and (\ref{newpartition}). 

Provided $V$ is sufficiently close to the minimum of a $E_i(\nu)$ curve at $V=V_i$ in Fig.~\ref{levels}, one can then use a parabolic approximation: 
\begin{equation}\label{parabolic}
E_i(\nu)=E_{i,0}+\frac{1}{2}K_i(\nu-\nu_i)^2/k_{\rm B}N,
\end{equation} 
where $K_i$ is the bulk modulus of the configuration and $\nu_i=(V_i-V_0)/V_0$ is the relative volume dilation at which it has its lowest energy. In this case, $\nu_i^\ast=\nu_i\big(\frac{K_i}{K_0}\big)$. It is important to emphasize, however, that the dilation parameter $\nu^\ast_i$ obtained from fitting is {\it not} the actual dilation associated with the equilibrium volume of a given valence state, but, rather, a renormalized dilation parameter, which limits our ability to make accurate estimates of the equilibrium volume of each of the excited valence states in $\delta$-Pu. However, this has no discernible impact on the calculations of the heat capacity and entropy.

When we compare the model against neutron scattering lattice parameter data in Fig.~\ref{data}d,\cite{lawson1} we use $\Delta a/a_0=\frac{1}{3}\nu(T,B=0)=\frac{1}{3}\big[\nu_{\rm ph}(T)+\nu_{\rm el}(T,B=0)\big]$, where $\Delta a$ refers to the fractional change in lattice parameter on increasing the temperature and $a_0$ refers to the value of the lattice parameter at zero temperature. The linear thermal expansion coefficient in Fig.~\ref{data}e is obtained by numerically differentiating the fitted form using $\alpha_l=(\partial a/\partial T)/a_0$. In the case of the magnetostriction, we neglect the phonon contribution and compare $s_\|\approx\frac{1}{\eta}\big[\nu_{\rm el}(T,B=B_{\rm max})-\nu_{\rm el}(T,B=0)\big]/B^2_{\rm max}$ against the quadratic coefficient of magnetostriction plotted in Fig.~\ref{data}c. Since $\delta$-Pu is both cubic and polycrystalline, $1\lesssim\eta\lesssim3$.\cite{haffner3} For the purpose of fitting to the longitudinal magnetostriction and volume expansion, we assume $\eta=$~3, although the actual value has no direct impact on the heat capacity entropy calculation. In our fits, $B_{\rm max}$ is the maximum magnetic field used in the magnetostriction experiments. The leading order quadratic form of the magnetostriction generally arises from the cancellation of the odd terms in the partition function upon summing the spin up and down pseudospin components. On substituting different values of $B_{\rm max}$ in the magnetostriction coefficient numerical simulations, we find no significant deviation from a conventional quadratic form in the model. 

\subsection{Fitting results} 
A least squares fit is performed simultaneously to both the quadratic magnetostriction coefficient and the thermal expansion volume of $\delta$-Pu$_{1-x}$Ga$_x$, with the minimization being made with respect to the product of the sum of the squares of both quantities. For the ground state, $\nu^\ast_0$, $E^\ast_0$ and $\mu^\ast_0$ are set to zero. On leaving $\mu^\ast_0$ as a free parameter, it goes to zero on performing the least squares fit. For $\nu^\ast_1$, $E^\ast_1$ and $\mu^\ast_1$ and $\nu^\ast_2$, $E^\ast_2$ and $\mu^\ast_2$, we fit an individual set of parameters for $x=$~2\% Ga and $x=$~7\% Ga. For $x=$~4\% and 6\%, the parameters used for calculating the volume are linearly interpolated between those at $x=$~2\% Ga and $x=$~7\% Ga, while for $x=$~0\%, an extrapolation is made. To minimize the number of parameters, $T_{\rm fl}$, $\theta_{\rm D}$ and $\gamma$ are assumed to be independent of $x$. During fitting, the valence fluctuation temperature $T_{\rm fl}$ is left as a free parameter. $T_{\rm fl}=$~0~K is therefore the best fit value of the valence fluctuation temperature (or degree of interconfigurational mixing). All fitted parameters are listed in Table~\ref{table1}. As a demonstration of self consistency of the fitted model, the activation energy $E^\ast_2$ of the upper excited electronic configuration and the magnitude of the change in its characteristic volume are both found to increase on reducing the amount of Ga (see Table~\ref{table1}). Conversely, the activation energy $E^\ast_1$ of the lower excited valence configuration and the magnitude of the change in its characteristic volume are both found to decrease on reducing the amount of Ga. 

\begin{table}[ht]
\caption{Values of the various parameters obtained on performing a least squares fit. The corresponding parameters of the ground state of $E_0^\ast=\mu^\ast_0=0$, while $V_0=$~24.57 and 23.87~\AA$^3$ for $x=$~2 and 7\%, respectively (estimated from diffraction measurements).\cite{lawson1}}
\centering
\begin{tabular}{c c c c c}\\
\hline\hline
Quantity & $x=$~2\% Ga & $x=$~7\% Ga & all $x$ & Units \\ [0.5ex] 
\hline
$(1+\nu_1^\ast)V_0$&24.66$\pm$~0.02&24.37~$\pm$~0.11&&\AA$^3$  \\
$E_1^\ast$&265~$\pm$~10&476~$\pm$~18&& K \\
$\mu_1^\ast$&1.7~$\pm$~0.3&1.3~$\pm$~0.3&& $\mu_{\rm B}$ \\
$(1+\nu_2^\ast)V_0$&21.29~$\pm$~0.01&23.71~$\pm$~0.08&&\AA$^3$  \\
$E_2^\ast$&1450~$\pm$~90&810~$\pm$~80&& K \\
$\mu_2^\ast$&2.9~$\pm$~1.0&3.8~$\pm$~1.4&& $\mu_{\rm B}$ \\
$T_{\rm fl}$ & & & 0~($+$~50)& K \\
$\gamma$ &  &  &0.50~$\pm$~0.04& -- \\ [1ex]
\hline
\end{tabular}
\label{table1}
\end{table}

\subsection{Heat capacity measurements}
\begin{figure}[!!!!!!!htbp]
\begin{center}
\includegraphics[angle=0,width=1\linewidth]{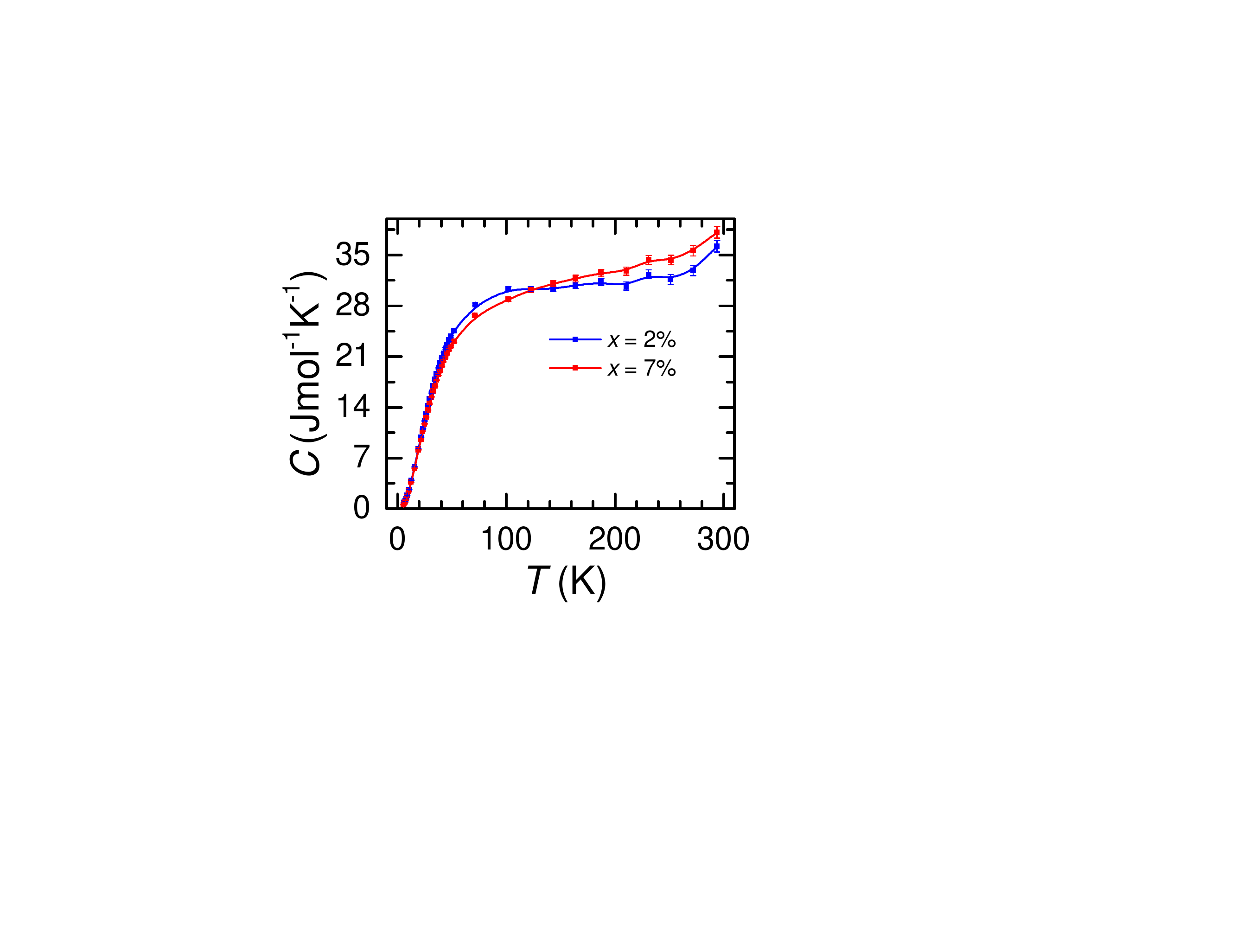}
\textsf{\caption{{\bf Heat capacity for $\delta$-Pu$_{1-x}$Ga$_x$ samples, as indicated}. 
}
\label{tobash}}
\end{center}
\end{figure}

Heat capacity measurements (see Fig.~\ref{tobash}) are made on samples of $\delta$-Pu$_{1-x}$Ga$_x$ (with $x=$~2\% and 7\%), whose masses are $\approx$~1~mg to minimize the effect of self heating under vacuum caused by self irradiation. The measurements are made in a standard Quantum Design physical properties measurement system (PPMS). While use of small samples increases the error associated with the subtraction of the addendum, but does not significantly impact the difference in heat capacity between 2\% and 7\% samples shown in Fig.~\ref{entropy}d. 

The difference in Pu content in each of the samples introduces a systematic error in the difference, and the extent to which this difference can be attributed to the electronic contribution. Since the contribution to the heat capacity from phonons universally saturates at the Dulong-Petit value of $\approx$~25~Jmol$^{-1}$K$^{-1}$ regardless of the Pu content, the subtracted quantity in Fig.~\ref{entropy}d is free from any significant phonon contribution above $\approx$~50~K. The accuracy of the remaining $\approx$~5~Jmol$^{-1}$K$^{-1}$ electronic contribution in each sample is affected by $\Delta x=$~5\% difference in Pu content, therefore making the systematic error in making the subtraction 5\% $\times$ 5~Jmol$^{-1}$K$^{-1}=$~0.25~Jmol$^{-1}$K$^{-1}$. However, this is significantly less than the measurement error bar in Fig.~\ref{entropy}d.

\subsection{Possible relation of fitting results to the electronic structure}
The observed changes in magnetostriction with increasing temperature indicate that the magnetic moment appears to be smallest for the ground state configuration (see Methods), suggesting its possible correspondence to $n_f=$~4 or 5 $5f$-electrons confined to the atomic core\cite{eriksson1,svane1,booth1}. Both of these configurations have the potential for orbital compensation to produce small moments\cite{wills1,soderlind3} compatible with the absence of magnetic ordering.\cite{lashley1} Partial occupancy of both $n_f=$~4 and 5 has further been suggested on the basis of neutron scattering structure factor measurements,\cite{janoschek1} although such measurements are performed at a temperature of $T=$~293~K that is sufficiently high for both to be thermally occupied. 

On incorporating valence fluctuations phenomenologically into $F(T,B,x)$, using the established methodology of an effective valence fluctuation temperature $T_{\rm fl}$ (see below),\cite{zieglowski1,wohlleben1} we find that $T_{\rm fl}\lesssim$~50~K, which is more than an order of magnitude smaller than the $\sim$~10$^3$~K suggested by existing refined electronic structure models.\cite{savrasov1,shim1,zhu1} 

\subsection{Energy level schematics}
\begin{figure}[!!!!!!!htbp]
\begin{center}
\includegraphics[angle=0,width=0.9\linewidth]{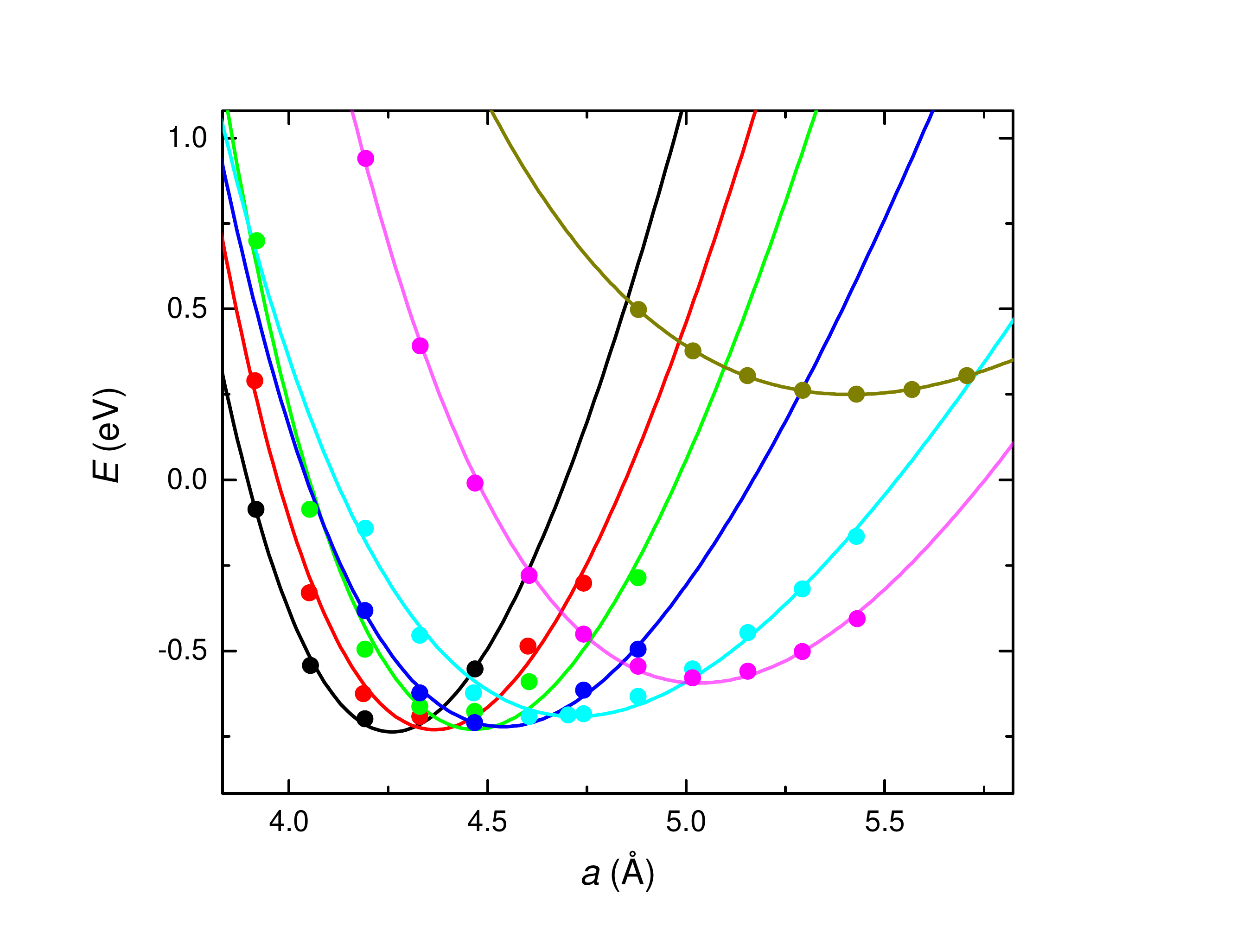}
\textsf{\caption{{\bf Configurational energies versus volume according to Svane {\it et al.}\cite{svane1}}. 
Symbols are the energy $E$ versus lattice parameter $a$ calculated according to Svane {\it et al.}\cite{svane1}, while lines are fits to Equation (\ref{cohesion}).
}
\label{svane}}
\end{center}
\end{figure}

Cohesion in metals is generally expected to give rise to an energy $E$ versus linear dimension of the form\cite{ashcroft1}
\begin{equation}\label{cohesion}
E=\bigg(a_0-\frac{a_1}{a}+\frac{a_2}{a^2}\bigg),
\end{equation}
where, here, $a=(4V)^{\frac{1}{3}}$ refers to the lattice parameter and $a_0$, $a_1$ and $a_2$ are constants. For the schematics in Figs.~\ref{levels}a, b and c, the $E$ versus $V$ curves are assumed to have this form. In Fig.~\ref{svane} we perform a fit of Equation (\ref{cohesion}) to the calculated energy for $\delta$-Pu of Svane {\it et al.}\cite{svane1}, confirming that Equation (\ref{cohesion}) is approximately valid for electronic structure calculations of plutonium. The minima occur at $a_{\rm min}=\frac{2a_2}{a_1}$, with the bulk modulus at $a=a_{\rm min}$ being given by $K=\frac{\partial^2E}{\partial\nu^2}_{a=a_{\rm min}}=\frac{1}{72N}({a_1^8}/{a_2^7})$. In Fig.~\ref{levels}c, the $E$ versus $V$ schematic has been calculated using Equation (\ref{cohesion}) so that $E_2-E_0$ and $E_1-E_0$ are consistent with $E^\ast_1$ and $E^\ast_2$ in Fig.~\ref{levels}d, respectively. Only the bulk modulus $K_0$ (at $a=a_{\rm min}$) associated with $E_0$ has been measured directly. For $E_1$ and $E_2$, we have arbitrarily assumed $K_1=$~40~GPa and $K_2=$~50~GPa for $E_1$ and $E_2$, respectively. 

The rapid fall of $E^\ast_2$ with decreasing volume suggests its minimum is located at a volume and energy that is significantly lower than that of $E_0$, which is consistent with $E_0$ being representative of a metastable configuration separated from a lower energy configuration by an energy barrier.\cite{hecker2,sadigh1} The volumes of $\alpha$ and $\alpha^\prime$ are similar to the equilibrium volume of $E_2$ in Fig.~\ref{levels}c, suggesting that the structural transformation to $\alpha$ may be a secondary effect associated with the volume collapse. For the volume collapse to occur, the net energy gain associated with the transition needs to be equal to or greater than the energy losses associated with what are essentially displacive structural transitions from $\delta$ to $\alpha$. In pure Pu$_{1-x}$Ga$_x$, the transitions from $\delta$ to $\alpha$ always occur for $x\lesssim$~2\%. However, for $x\gtrsim$~2\%, the $\delta$-phase continues to persist down to low temperatures. While we have identified entropy to be an important factor, the insolubility of Ga in $\alpha$-Pu$_{1-x}$Ga$_x$ for large $x$ is another important factor. It has been shown experimentally that $\alpha$-Pu$_{1-x}$Ga$_x$ decomposes into $\alpha$-Pu and Pu$_3$Ga over timescales of order 10,000 years.\cite{hecker2,sadigh1}

\subsection{Comparison to the Invar model thermodynamic treatment}
Magnetostriction measurements are necessary for accurately separating the electronic and phonon contributions to the lattice, causing prior efforts to separate these contributions in the absence of magnetostriction data to be of limited success.\cite{lawson1} Because the upper excitation energy $E^\ast_2$ leads to a negative thermal expansion, which is obviously quite distinct from the Debye function, this higher excitation energy was successfully extracted in prior studies. However, because the positive thermal expansion associated with $E^\ast_1$ has the same sign as the thermal expansion caused by phonons, it was subsequently missed. In contrast to our three level model, the two level Invar model was shown to be unable to account for the excess entropy and heat capacity measured in $\delta$-Pu.\cite{lawson1}

Significant changes to the free energy partition function in the present approach include (1) the addition of effective magnetic moments associated with each configuration, (2) the addition of a third level, which is necessary for accurately reproducing the forms of the magnetostriction and heat capacity, and (3)  a more realistic modeling of the volume-dependences of the various configurations, which are assumed to have well defined minima in accordance with electronic structure calculations.\cite{eriksson1,svane1} The latter approach enables the bulk modulus to be derived from the partition function, which was not possible in in the Invar model without arbitrarily adding an extra term to the free energy.\cite{lawson1}

An underlying weakness of the Invar model,\cite{lawson1} is that atomic sites with different configurations were assumed to have different volumes, for which there is no evidence in $\delta$-Pu$_{1-x}$Ga$_x$.\cite{migliori1} In the present approach, by contrast, the volume $V$ is the same for each of the configuration. Only the dependence of $E_i$ on volume are assumed to be different.  

\subsection{Collapse of the bulk modulus with increasing temperature}
A well known observation in Ga-stabilized $\delta$-Pu is the strong reduction in the bulk modulus with increasing temperature, which occurs in regimes in which the thermal expansion is both positive and negative.\cite{migliori1} In the Invar model, this was attributed to the lower volume excited state having a bulk modulus close to zero. In the present multiconfigurational approach, the negative thermal expansion is a natural consequence of having an excited configuration whose equilibrium volume is substantially lower that $V_0$, and also generally asymmetric form of $E_i$ with respect to a change in the volume. To leading order, the bulk modulus is given by
\begin{equation}\label{bulkmodulus}
K=\frac{\partial^2F}{\partial\nu^2}\bigg|_T\approx\frac{1}{{Z_{\rm el}}^\ast}\sum_{i=0,1,2}\sum_{\sigma=\pm\frac{1}{2}} Nk_{\rm B}\frac{\partial^2E_i}{\partial\nu^2}e^{-\frac{k_{\rm B}E_i^\ast+2\sigma\mu_i^\ast B}{k_{\rm B}T^\prime}}.
\end{equation}
For configurations in which Equation (\ref{parabolic}) is a good approximation, $\frac{\partial^2E_i}{\partial\nu^2}=K_i/Nk_{\rm B}$, in which case its contribution to the bulk modulus becomes $\frac{1}{{Z_{\rm el}}^\ast}\sum_{\sigma=\pm\frac{1}{2}}K_ie^{-\frac{k_{\rm B}E_i^\ast+2\sigma\mu_i^\ast B}{k_{\rm B}T^\prime}}$. More generally, however, $E_i(\nu)$ is asymmetric about $V_i$ [see for example Equation (\ref{cohesion}) and Fig.~\ref{levels}c],\cite{eriksson1,svane1} having the potential to cause $\frac{\partial^2E_i}{\partial\nu^2}$ to depart significantly from $K_i/Nk_{\rm B}$. 
When excitations occur to a state for which $V$ is less than its equilibrium value $V_i$, $\frac{\partial^2E_i}{\partial\nu^2}=V^2\frac{\partial^2E_i}{\partial V^2}$ climbs steeply with decreasing $V$ (e.g. the case of $E^\ast_1$ in Fig.~\ref{levels}c). However, when excitations occur to a state for which $V$ is greater than its equilibrium value, $\frac{\partial^2E_i}{\partial\nu^2}=V^2\frac{\partial^2E_i}{\partial V^2}$ falls with increasing $V$, and may even turn negative (see e.g. the case of $E^\ast_2$ in Fig.~\ref{levels}c). 

Since a zero or negative $\frac{\partial^2E_i}{\partial\nu^2}$ term in Equation (\ref{bulkmodulus}) generally occurs only for excited configurations whose equilibrium volume dilations $V_i$ smaller than $V_0$, the experimentally observed\cite{migliori1} rapid decrease of the bulk modulus of Ga-stabilized $\delta$-Pu with increasing temperature could constitute further supporting evidence for an excited configuration with a small equilibrium volume $V_i$.

\subsection{Expanding fitting to include the magnetic susceptibility}
While the magnetostriction and thermal expansion measurements indicate that Ga-stabilized $\delta$-Pu most likely settle into a non-magnetic or weakly magnetic configuration at low temperatures, heat capacity and magnetic susceptibility measurements indicate the coexistence of a Fermi liquid state, in which both the Sommerfeld coefficient\cite{lashley2} and Pauli susceptibility\cite{lashley1} are enhanced. In the mixed level picture, in which the ground state configuration consists of $n_f=$~4 $5f$-electrons confined to the atomic core, the Fermi liquid state is predicted to originate from the one $5f$-electron that is itinerant, and its hybridization with other states.\cite{wills1} If the ground state configuration consists, instead, of that with $n_f=$~5 $5f$-electrons confined to the atomic core, then a Fermi liquid state could result from their weak hybridization with other states. 

\begin{figure}[!!!!!!!htbp]
\begin{center}
\includegraphics[angle=0,width=0.9\linewidth]{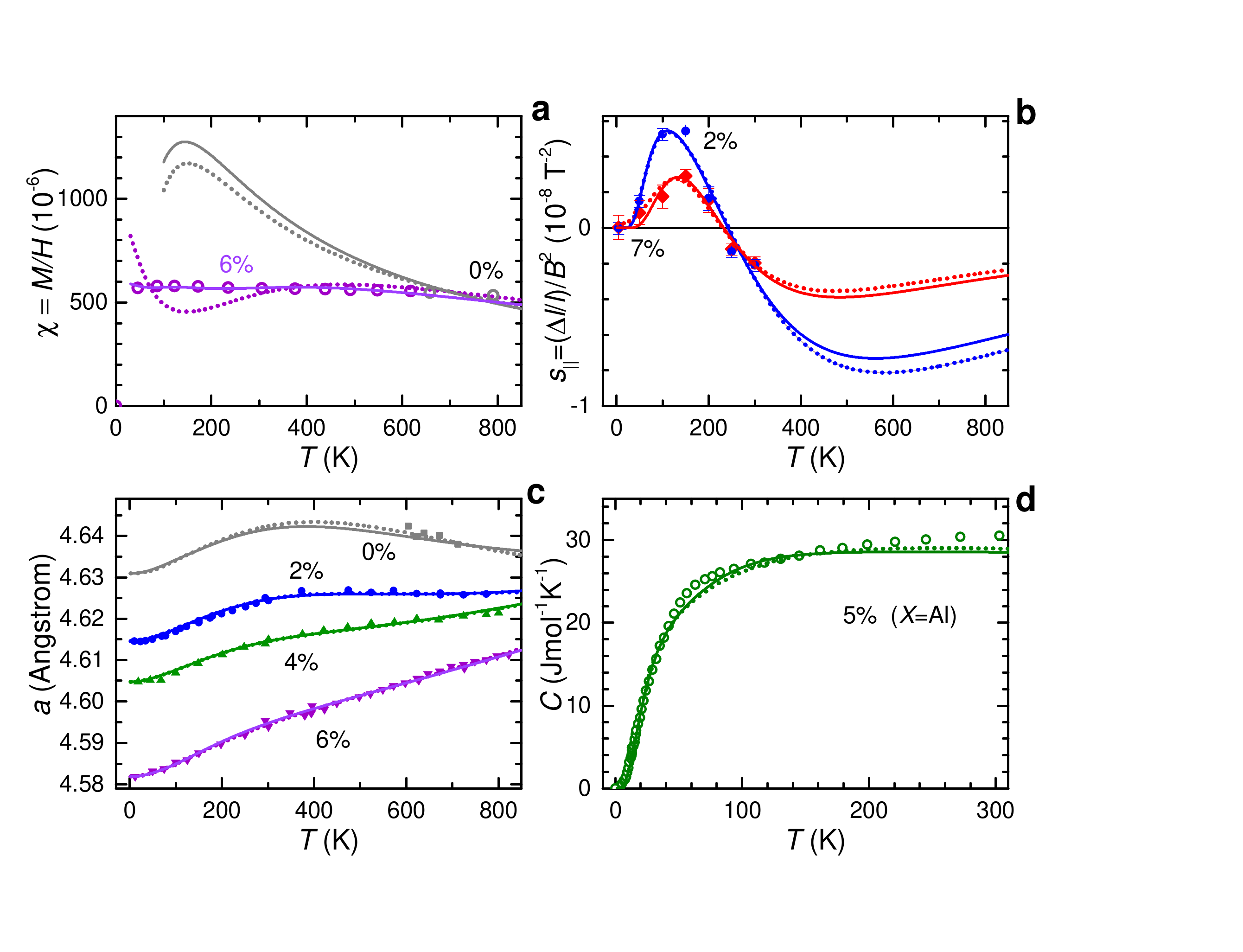}
\textsf{\caption{{\bf Results of fitting expanded to include the susceptibility}. {\bf a} Magnetic susceptibility of $\delta$-Pu$_{1-x}$Ga$_x$ versus $T$ for $x=$~0\% (pure Pu) and $x=$~6\%. {\bf b} Magnetostriction coefficient of $\delta$-Pu$_{1-x}$Ga$_x$ versus $T$ for $x=$~2\% and $x=$~7\%. {\bf c} Lattice parameter of $\delta$-Pu$_{1-x}$Ga$_x$ versus $T$ for $x=$~0\%, 2\%, 4\% and $x=$~6\%. {\bf d} Heat capacity of $\delta$-Pu$_{1-x}$Al$_x$ versus $T$ for $x=$~5\%. Data points represent experimental data referred to in the main text, while lines refer to fits. The fits are made to the susceptibility, magnetostriction coefficient and lattice parameter using free energy functional either with $T_{\rm fl}$ included as a finite parameter (dotted lines) or by inserting an alternative form for the susceptibility (solid lines) that better reproduces a Fermi gas-like form for the ground state configuration (see Methods). No fit is made to the heat capacity. In this case the lines are simply calculated from the free energy and compared against the measured curve.
}
\label{suscfit}}
\end{center}
\end{figure}

Any attempt to model the magnetic susceptibility requires additional fitting parameters to be introduced and is ultimately limited by the relative scarcity of available experimental data as a function of both $T$ and $x$.\cite{lashley1,meotraymond1} One approach is to utilize the methodology in which a finite $T_{\rm fl}$ mimics the behavior of a Fermi liquid at low temperatures.\cite{zieglowski1,wohlleben1} In this case, both $T_{\rm fl}$ and the magnetic moment of the ground state configuration, $\mu^\ast_0$, must acquire finite values. In order to reconcile the magnitude of the magnetic moment inferred from longitudinal magnetostriction and magnetic susceptibility measurements, $\eta$ also needs to be considered as an adjustable parameter. The magnetic susceptibility is given by $\chi_{zz}=\mu_0\partial M_{zz}/\partial B$, where
\begin{equation}\label{electronicsusc1}
M_{zz}(T,B)=-\frac{\partial F_{\rm el}}{\partial B}=\frac{1}{{Z_{\rm el}}^\ast}\sum_{i=0,1,2} 2\mu_i^\ast e^{-\frac{k_{\rm B}E_i^\ast}{k_{\rm B}T^\prime}}\sinh\bigg(\frac{\mu_i^\ast B}{k_{\rm B}T^\prime}\bigg).
\end{equation}
The results of a combined fit to the magnetostriction, volume expansion and magnetic susceptibility are shown in Table~\ref{table2} and Fig.~\ref{suscfit}.

\begin{table}[ht]
\caption{Values of the various parameters obtained on performing a least squares fit including the susceptibility. Again, $E_0^\ast=0$.}
\centering
\begin{tabular}{c c c c c}\\
\hline\hline
Quantity & $x=$~2\% Ga & $x=$~7\% Ga & all $x$ & Units \\ [0.5ex] 
\hline
$\mu_0^\ast$&0.0~$\pm$~0.1&0.5~$\pm$~0.2&& $\mu_{\rm B}$ \\
$(1+\nu_1^\ast)V_0$&24.64~$\pm$~0.02&24.26~$\pm$~0.11&&\AA$^3$  \\
$E_1^\ast$&275~$\pm$~10&458~$\pm$~18&& K \\
$\mu_1^\ast$&1.4~$\pm$~0.3&1.0~$\pm$~0.3&& $\mu_{\rm B}$ \\
$(1+\nu_2^\ast)V_0$&21.30~$\pm$~0.01&23.63~$\pm$~0.08&&\AA$^3$  \\
$E_2^\ast$&1360~$\pm$~90&890~$\pm$~80&& K \\
$\mu_2^\ast$&1.9~$\pm$~1.0&2.5~$\pm$~1.4&& $\mu_{\rm B}$ \\
$T_{\rm fl}$ &12~$\pm$~50 & 58~$\pm$~50 &  & K \\
$\gamma$ &  &  &0.54~$\pm$~0.04& -- \\
$\eta$&1.6~$\pm$~0.4&1.4~$\pm$~0.4&& -- \\[1ex]
\hline
\end{tabular}
\label{table2}
\end{table}

An alternative approach is to substitute the $\sinh(\mu_i^\ast B/k_{\rm B}T^\prime)$ term for the ground state configuration in Equation~(\ref{electronicsusc1}) with a Fermi gas-like form
\begin{equation}\label{fermiliquidform}
{\rm f}(E,B,W,\mu_{\rm F}^\ast)=\frac{2}{\sqrt{\pi}W}\int^\infty_{-\infty}\sum_{\sigma=\pm\frac{1}{2}} 2\sigma e^{-\big(\frac{E+2\sigma\mu_{\rm F}^\ast B}{W}\big)^2{\rm f}_{\rm FD}(E,T)}{\rm d}E,
\end{equation}
where we have assumed a Gaussian line shape for the electronic density-of-states, given the unknown band topology. Here, ${\rm f}_{\rm FD}=(1+e^{\frac{E}{T}})^{-1}$ is the Fermi-Dirac distribution function while $W$ is the electronic bandwidth. The results of a combined fit to the magnetostriction, volume expansion and magnetic susceptibility are shown in Table~\ref{table3} and again in Fig.~\ref{suscfit}.

\begin{table}[ht]
\caption{Values of the various parameters obtained on performing a least squares fit including the susceptibility. Again, $E_0^\ast=0$, and we have also set $T_{\rm fl}=0$.}
\centering
\begin{tabular}{c c c c c}\\
\hline\hline
Quantity & $x=$~2\% Ga & $x=$~7\% Ga & all $x$ & Units \\ [0.5ex] 
\hline
$(1+\nu_1^\ast)V_0$&24.64~$\pm$~0.02&24.21~$\pm$~0.11&&\AA$^3$  \\
$E_1^\ast$&271~$\pm$~10&388~$\pm$~18&& K \\
$\mu_1^\ast$&1.3~$\pm$~0.3&0.8~$\pm$~0.3&& $\mu_{\rm B}$ \\
$(1+\nu_2^\ast)V_0$&21.41~$\pm$~0.01&23.53~$\pm$~0.08&&\AA$^3$  \\
$E_2^\ast$&1320~$\pm$~90&970~$\pm$~80&& K \\
$\mu_2^\ast$&1.7~$\pm$~1.0&2.5~$\pm$~1.4&& $\mu_{\rm B}$ \\
$\gamma$ &  &  &0.55~$\pm$~0.04&-- \\
$W$& & & 408~$\pm$~20 & K \\
$\mu_{\rm F}^\ast$& & &0.8~$\pm$~0.2& $\mu_{\rm B}$ \\
$\eta$&1.3~$\pm$~0.4&1.8~$\pm$~0.4&& -- \\[1ex]
\hline
\end{tabular}
\label{table3}
\end{table}

While the former approach using $T_{\rm fl}\neq$~0 yields small magnetic moments for the ground state configuration, the latter approach in which the ground state configuration is accompanied by a half-filled electronic band more accurately reproduces the form of the magnetic susceptibility for Ga-stabilized $\delta$-Pu with $x=$~6\% as a function of temperature.\cite{meotraymond1} With both approaches, the respective energy levels $E^\ast_1$ and $E^\ast_2$ for $x=$~2\% and 7\% change very little on including the magnetic susceptibility. A general prediction of both approaches is that the temperature-dependence of the magnetic susceptibility becomes stronger on reducing $x$, which is not too surprising given that volume magnetostriction is proportional to the volume-dependence of the magnetization.  The development of a more refined model will require comprehensive measurements of the susceptibility as a function of both $T$ and $x$.

\clearpage

\begin{figure}[!!!!!!!htbp]
\begin{center}
\includegraphics[angle=0,angle=0,width=0.9\linewidth]{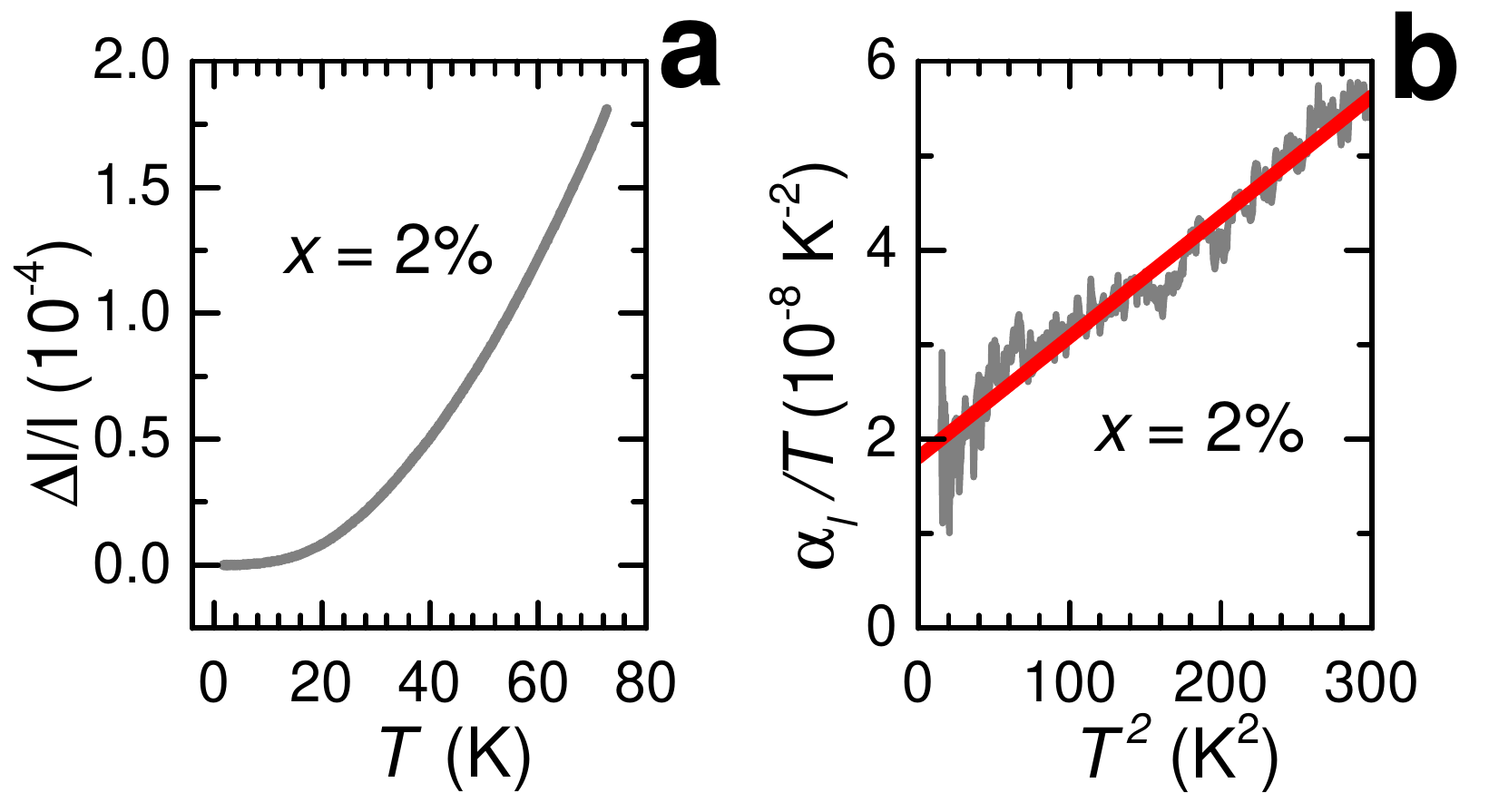}
\textsf{\caption{{\bf Low temperature thermal expansion}. {\bf a} Change in length $\Delta l/l$ at low $T$ in $\delta$-Pu$_{1-x}$Ga$_x$ ($x=$~2\%) obtained using the optical fiber Bragg grating method. {\bf b} Plot of $\alpha_l/T$ versus $T^2$ of $\delta$-Pu$_{1-x}$Ga$_x$ ($x=$~2\%) at low $T$ and $x=$~2\%, together with a linear fit (red line) to the function $\alpha_l/T=\frac{\kappa_0}{3}(k_{\rm ph}T^2+\frac{2}{3}\gamma_{\rm el})$, where $k_{\rm ph}$ is a constant relating to phonons and $\gamma_{\rm el}$ is the linear-in-$T$ electronic contribution to the heat capacity.\cite{ashcroft1} According to the fit, $\gamma_{\rm el}=$~(40~$\pm$~10)~mJmol$^{-1}$K$^{-2}$, which is of comparable order to the result ($\gamma_{\rm el}=$~64~mJmol$^{-1}$K$^{-2}$) obtained for $\delta$-Pu$_{1-x}$Al$_x$ ($x=$~5\%) from heat capacity measurements.\cite{lashley2} 
}
\label{lowTthermal}}
\end{center}
\end{figure}

\begin{figure}[!!!!!!!htbp]
\begin{center}
\includegraphics[angle=0,angle=0,width=0.45\linewidth]{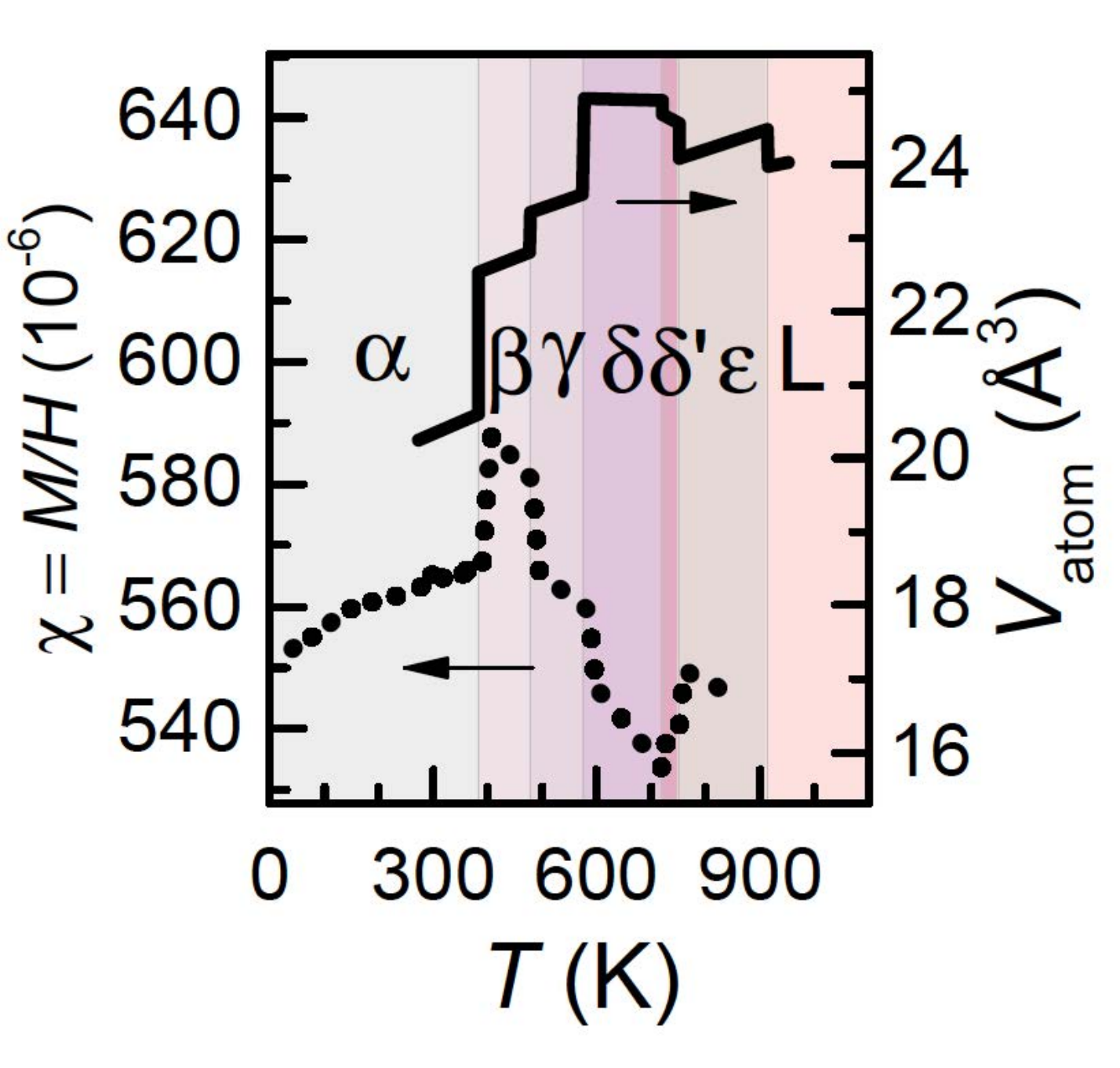}
\textsf{\caption{{\bf Magnetovolume effect in Pu}. Plots of the atomic volume and susceptibility of Pu versus temperature, with the different crystalline phases shaded in different colors. 
}
\label{extrastuff}}
\end{center}
\end{figure}

%

\end{document}